\documentclass[aps,pre,amsmath,amssymb,reprint,floatfix]{revtex4-1}
\makeatletter
\newcommand*{\rom}[1]{\expandafter\@slowromancap\romannumeral #1@}
\makeatother
\usepackage{lipsum}
\usepackage{amssymb,euscript,latexsym,amsmath,amsthm,subeqnarray}
\usepackage{graphicx}
\usepackage{color}
\usepackage{setspace}
\usepackage{float}
\usepackage{bm}
\usepackage[dvipsnames]{xcolor}
\usepackage{subcaption}
\usepackage{cancel}
\usepackage{setspace}
\usepackage{graphicx}
\usepackage{siunitx}

\usepackage[T1]{fontenc}
\usepackage[utf8]{inputenc}
\usepackage[font=small,labelfont=bf,tableposition=top]{caption}
\usepackage{booktabs,multirow}

\DeclareGraphicsExtensions{.eps,.ps,.pdf}

\def\f#1#2{{\frac{#1}{#2}}}
\def\*{\cdot}

\def\t{\mathbf{t}}

\def\f{\mathbf{f}}
\def\r{\mathbf{r}}

\def\v{\mathbf{v}}
\def\I{\mathbf{I}}

\def\dy{\delta y}

\def\dy{\delta y }

\def\h{{\rm h}}

\def\re{{\rm e}}

\def\ua{^{(1)}}
\def\ub{^{(2)}}

\def\ui{^{(i)}}
\def\uj{^{(j)}}
\def\uji{^{(j\to i)}}

\begin{document}

\title{Multisynchrony in active microfilaments}

\author{Yi Man}
\author{Eva Kanso}
\email{kanso@usc.edu}
\affiliation{Department of Aerospace and Mechanical Engineering,  \\ University of Southern California, Los Angeles, California 90089, USA}

\date{\today}
\begin{abstract}
Biological microfilaments exhibit a variety of synchronization modes. Recent experiments observed that a pair of isolated eukaryotic flagella, coupled solely via the fluid medium, display synchrony at nontrivial phase-lags in addition to in-phase and anti-phase synchrony. Using an elasto-hydrodynamic filament model in conjunction with numerical simulations and a Floquet-type theoretical analysis, we demonstrate that it is possible to reach multiple synchronization states by varying the intrinsic activity of the filament and the strength of hydrodynamic coupling between the two filaments. We then derive an evolution equation for the phase difference between the two filaments at weak-coupling, and use a Kuramoto-style phase sensitivity analysis to reveal the nature of the bifurcations underlying the transitions between these different synchronized states. 
\end{abstract}

\maketitle

Biological microfilaments, such as cilia and flagella, exhibit a variety of synchronization modes. 
As the surrounding fluid is an obvious medium for force transmission, hydrodynamic interactions are deemed crucial for synchronization. 
Two flagella isolated from the somatic cells of \textit{Volvox carteri}, and later of \textit{Chlamydomonas reinhardtii}, and thus coupled via the fluid medium only, synchronize their beating in-phase or anti-phase at close interflagellar distance \cite{Brumley2014,Wan2014}. Synchronized states with nontrivial phase lags, between 0 and $\pi$,  have also been observed, but not thoroughly analyzed \cite{Wan2014}.

Taylor pioneered the theoretical study of fluid-mediated synchronization by considering two infinite sheets with prescribed traveling waves; he found that in-phase synchronization is stable and exhibits minimum viscous energy dissipation \cite{Taylor1951}. Later,  anti-phase synchrony, characterized by maximum dissipation, was also shown to be stable \cite{Elfring2009}.  Synchronization was also analyzed in experiments with driven colloids~\cite{Kotar2010,Bruot2012, Bruot2016} and in far-field models~\cite{Vilfan2006,Niedermayer2008,Uchida2011, Uchida2012}, assuming that the interfilamentous distance $h$ is much larger than the filament length $L$ so that each filament can be modeled as an oscillating  bead.  Due to time-reversibility of the Stokes equations,  
in addition to hydrodynamic coupling, either a non-constant force profile \cite{Uchida2011, Uchida2012} or orbital compliance \cite{Niedermayer2008} is necessary to achieve synchrony in the bead model. The synchronized state depends on the force profile and  shape of the orbit. Particularly, for circular orbits, only in-phase synchrony is stable, while for select elliptic orbits, synchronized states with opposite phase or nontrivial phase lag appear to be stable \cite{Vilfan2006, Uchida2011, Uchida2012}. 
Despite the richness of these weakly-coupled bead models, in most biological situations, the opposite regime where $h\ll L$ is more relevant and the slender geometry of the filament should be considered. Until recently, there are a few elastic filament models for synchronizations, and they mainly focus on in-phase and anti-phase synchronizations only \cite{Goldstein2016, Guo2018, Kawamura2018, Chakrabarti2019,SteinArxiv}.

\begin{figure}[!t]
	\centering
	\includegraphics[scale=1]{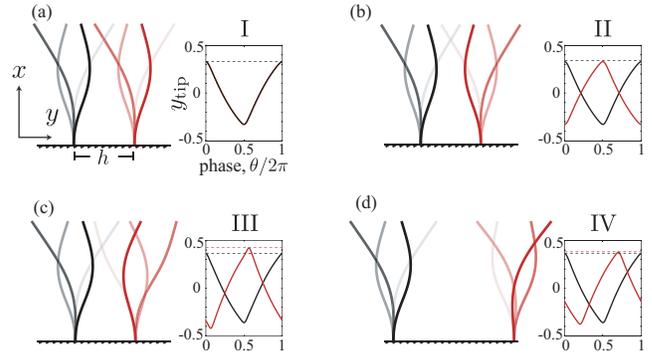}
	\caption{\footnotesize  Two active microfilaments synchronize their beating: (a) in-phase for $F = 46$, $\gamma = 0.1$, $\Delta\theta_0/2\pi = 0.2$; (b) anti-phase for $F = 46$, $\gamma = 0.1$, $\Delta\theta_0/2\pi = 0.3$; (c) at different amplitudes for $F = 49$, $\gamma = 0.1$, $\Delta\theta_0/2\pi = 0.3$; (d) at a nontrivial phase lag ($\Delta \theta \approx 0.56\pi$) for $F = 48$, $\gamma = 0.01$, $\Delta\theta_0/2\pi = 0.3$. In each panel, left plots depict snapshots of the steady-state waveforms, with increasing time highlighted in darker color, and right plots show the tip deflection over one period of oscillation as a function of phase $\theta/2\pi$. These four cases, labeled \rom{1}-\rom{4}, are highlighted in Fig.~\ref{fig:basins}  \&\ref{fig:synctime}.		}
	\label{fig:modes}
\end{figure}

Biological microfilaments, namely cilia and eukaryotic flagella, are driven into sustained oscillations by an intricate internal structure of microtubule doublets and dynein motors~\cite{Nicastro2006, Lin2018}. The spatial and temporal regulations of this molecular machinery are still under debate; see, e.g.,~\cite{Bayly2016, Sartori2016b, Han2018, Man2019} and references therein. 
We posit that the exact details driving flagellar oscillations matter little to the coordination of multiple flagella. In this Letter, we apply a recently-proposed phenomenological model, in which the active motor forces are represented by a tangential force  $F$ exerted at the filament tip \cite{DeCanio2017, Ling2018,DeCanioThesis}. We show that two filaments coupled via near-field hydrodynamics ($h\ll L$) can reach multiple synchronized states with the same, opposite, and even nontrivial phase lags.  We analyze the stability and basins of attraction of these states using Floquet theory and Kuramoto-style phase reduction analysis.


In particular, we consider two identical filaments of radius $a$ and length $L\gg a$, clamped at their base at a distance $h$ apart, and subject to an applied tangential force $F$ at their tip. 
We let $\r(s,t)$ denote the  position of one of the filaments as a function of time $t$ and arclength $s$. The subscripts $(\cdot)_t$ and $(\cdot)_s$ denote differentiation with respect to $t$ and $s$, respectively. 
The hydrodynamic force density $\f_\h$ is anisotropic and proportional to the filament velocity relative to the fluid velocity; namely,
$\f_\h = -\xi\left(\I-\frac{1}{2}\t\t\right)\cdot (\r_t-\v)$. Here, we introduced the unit tangent $\t = \r_s$ andt used the approximation that the perpendicular drag coefficient $\xi= 4\pi \mu/\ln(L/a)$ is twice as large as the tangential drag coefficient \cite{Lauga2009, Lighthill1976}. The vector $\v$ represents the fluid velocity induced by the motion of the other filament; it is identically zero for a single filament. For planar motions (in the $(x,y)$-plane), it is a classic result that the elastic force is given by $\f_\re = -B\r_{ssss}+(\Lambda\t)_s$, where $B$ denotes the bending rigidity and $\Lambda$ the tension enforcing filament inextensibility \cite{Wiggins1998}. Balance of forces $\f_{\h} + \f_{\re} = \mathbf{0}$ on each filament, together with clamped-free boundary conditions, lead a system of equations for the filaments dynamics. We express this system in non-dimensional form by choosing the length scale $L$ and the time scale given by the bending relaxation time $\xi L^4/B$; the local force scales as $B/L^3$. 

\begin{figure*}[!t]
	\centering
	\includegraphics[scale=1]{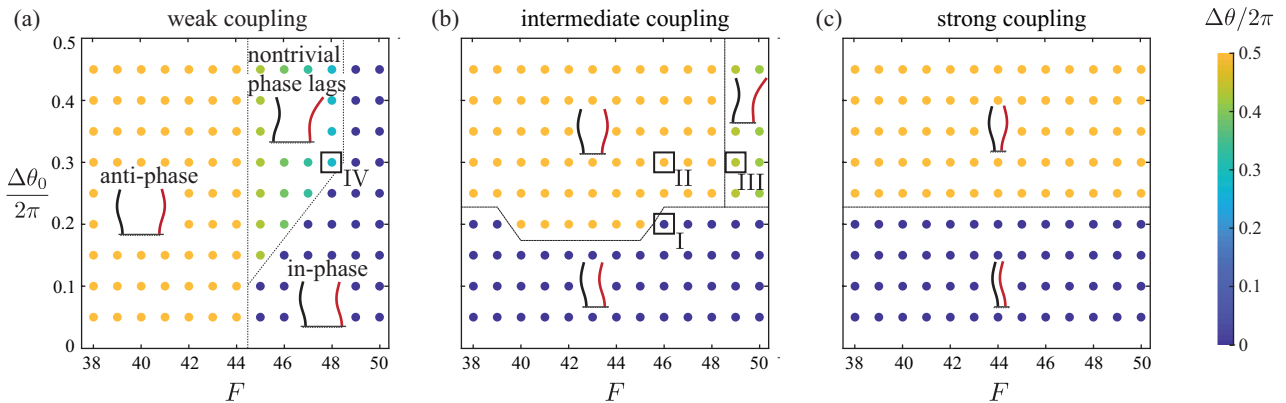}
	\caption{\footnotesize
		Basins of attraction of the synchronization modes: anti-phase, in-phase, and synchrony at non-trivial phase lag, as a function of initial phase difference $\Delta\theta_0$ and the active force value $F$. The phase difference between two filaments at steady state, $\Delta\theta$, is indicated by the color of the dots. We vary the inter-filament spacing such that (a) $\gamma = 0.01$, (b) $\gamma = 0.1$ and (c) $\gamma = 0.33$,  representing weak, intermediate and strong hydrodynamic couplings.  
	}
	\label{fig:basins}
\end{figure*}

To close the system, we use $\v\uji$ to denote the flow field at filament $i$ induced by the motion of filament $j$. For $a\ll L$ and $a\ll h$, the fluid velocity can be represented by that induced by a line of Stokeslets distributed along the centerline \cite{Hancock1953}. In the biologically relevant limit $a\ll h\ll L$, Man and co-authors calculated that, at the leading order, 
$\v\uji(s) = ({\ln(h(s)/L)}/{4\pi\mu})(\I+\t\t)\cdot\f_{\rm h}\uj(s)$, where $h(s)$ is the distance between two filaments at arc length $s$ \cite{Man2016, Man2017,ManThesis}. We assume that the wave amplitude is at most of the same order as the basal inter-filamentous distance $h$, and define $\gamma = \ln(L/h)/\ln(L/a)$ to indicate the strength of hydrodynamic coupling \cite{Goldstein2016}. The flow velocity reduces to $\v\uji(s) = -\gamma \r_t\uj(s)$.
As the displacement in the $y$-direction is dominant, we relax the inextensibility condition  and approximate $s\approx x$.
The equations governing the coupled oscillations of the two filaments are given by
\begin{equation}\label{eq:fbtwo} 
\begin{split}
y\ua_t	- \gamma y\ub_t & = -y\ua_{xxxx}+2\Lambda_x\ua y_{x}\ua+\Lambda\ua y_{xx}\ua,\\
y\ub_t	- \gamma y\ua_t & = -y\ub_{xxxx}+2\Lambda_x\ub y_{x}\ub+\Lambda\ub y_{xx}\ub,
\end{split}
\end{equation}
subject to the boundary conditions (for $i=1,2$)
$$\left.y^{(i)}\right|_{x=0}=\left.y^{(i)}_x \right|_{x=0}= \left.y_{xx}^{(i)}\right|_{x=1}=\left.y^{(i)}_{xxx} \right|_{x=1}=0.$$ 
We solve Eq.~\eqref{eq:fbtwo} numerically using an implicit finite difference scheme similar to that in \cite{Tornberg2004}; see SI \cite{SI}.

For a single filament, $\Lambda(x) = -F$ holds along the filament. As the active force $F$ exceeds a critical value $F_{\rm cr} = 37.5$, the filament buckles, and its linear dynamics is characterized by unstable oscillations with growing amplitude \cite{DeCanio2017, Ling2018}. To saturate the oscillation amplitude, we modify the tension by adding a nonlinear function of curvature, $\Lambda = -F+\alpha y_{xx}^2$, where the square comes from consideration of symmetry and $\alpha$ is a constant that we fix to $\alpha =4$. The beat frequency depends on $F$ not $\alpha$. The steady state behavior of the single filament follows a periodic, limit-cycle solution $y_0(x,t; F)$, such that $y_0(x,t+T;F) = y_0(x, T;F)$;  see SI and Fig.~S1 \cite{SI}. We define a phase parameter $\theta$ such that the filament configuration can be parameterized by its phase in the oscillation cycle  $y_0(s,\theta(t))$, and $\theta_t = \omega$, where $\omega/2\pi= 1 /T$ is the oscillation frequency.

For two coupled filaments, we identify two periodic solutions $y\ui(x,t;F,\gamma)$ by direct inspection of Eq.~\eqref{eq:fbtwo}: one solution $y\ua = y\ub = y_0(s,\theta)$, $\theta_t = (1-\gamma)^{-1}\omega$ corresponds to the two filaments synchronizing in-phase and following the same waveform as that of a single filament albeit at a higher frequency;  another solution, $y\ua = -y\ub = y_0(s,\theta)$, $\theta_t = (1+\gamma)^{-1}\omega$ corresponds to anti-phase synchrony at a lower frequency. The fact that in-phase solutions exhibit higher beat frequencies is consistent with recent experimental observations and mathematical models~\citep{Leptos2013,Wan2014,Guo2018}.



We initialize the two filaments at different phases  $y^{(i)}(x, t=0; F,\gamma) = y_0(x, \theta_0\ui)$ and solve Eq.~\eqref{eq:fbtwo} numerically. The steady state depends on the initial phase difference $\Delta\theta_0 = \theta_0\ub-\theta_0\ua$ and the parameter values $F$ and $\gamma$. Interestingly, as we vary $\Delta \theta_0$, $F$ and $\gamma$, we find synchronization modes other than the in-phase and anti-phase synchrony described above. In Fig.~\ref{fig:modes}, we show four examples labeled \rom{1} to \rom{4}. In \rom{1} and \rom{2}, the filaments converge to in- and anti-phase synchrony, respectively. In \rom{3}, the filaments oscillate nearly out-of-phase at different amplitudes while in \rom{4}, the amplitudes are almost identical, and the filaments synchronize at a non-trivial phase lag, $\Delta\theta/2\pi = 0.28$. 

An analysis of the net hydrodynamic forces on the coupled filaments in Fig.~\ref{fig:modes} (see SI and Fig.~S3 \& S4 \cite{SI}) shows that (i) compared to the single filament, the hydrodynamic force on each filament during in- and anti-phase synchrony remains the same, 
 (ii) asymmetric synchrony produces asymmetric forces that could result in a net moment on the filament pair, and (iii) the total force on both filaments is independent of the synchronization mode and coupling strength $\gamma$.
 
We next investigate the basins of attraction of these synchronization modes by systematically varying the initial phase difference $\Delta\theta_0$ for distinct values of $F$ and $\gamma$. In Fig.~\ref{fig:basins}, we plot the results on the $(\Delta\theta_0, F)$ space, for $\gamma = 0.01$, $0.1$ and $0.33$, which represent weak, intermediate, and strong hydrodynamic coupling, respectively. 
We observe that the dynamics strongly depends on $\gamma$. Under weak hydrodynamic coupling (Fig.~\ref{fig:basins}a), all three synchronization modes are observed. For $F \lesssim 44$, the filaments are always synchronized anti-phase. For $45\lesssim F \lesssim 48$, the filaments exhibit bistable behavior; they synchronize either in-phase or with nontrivial phase lags ranging approximately from $0.56\pi$ to $\pi$, as represented by the color bar on the far right. For $F\gtrsim 49$, only in-phase synchrony is observed. Under intermediate coupling (Fig.~\ref{fig:basins}b), the filaments exhibit bistable behavior for all $F$ with one transition: For $F\lesssim48$, the filaments synchronize either in-phase or anti-phase, while for $F\gtrsim 49$, they synchronize either in-phase or at a nontrivial phase lag.  For strong coupling (Fig.~\ref{fig:basins}b), the filaments exhibit bistability between in-phase and anti-phase synchrony.

\begin{figure}[!b]
	\centering
	\includegraphics[scale=1]{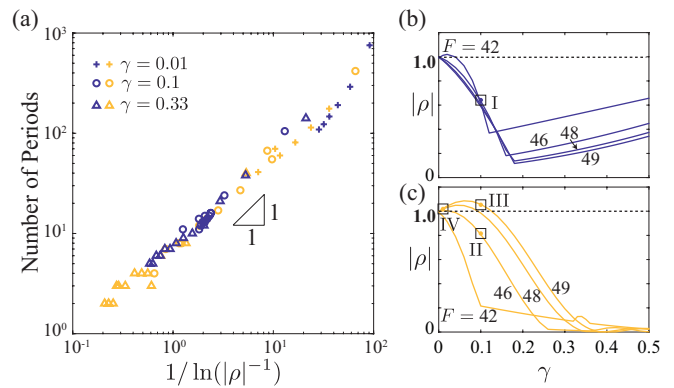}
	\caption{\footnotesize
	(a) Number of periods it takes to reach in-phase (blue) or anti-phase (yellow) synchrony versus $1/\ln(|\rho|^{-1})$; symbols represent distinct values of $\gamma$ and $F$. Floquet multipliers versus $\gamma$ for $F =42$, 46, 48 and 49, corresponding to (b) in-phase and (c) anti-phase synchrony.}
	\label{fig:synctime}
\end{figure}

We analyze the stability of in-phase and anti-phase synchrony using Floquet theory.
Considering first the case of in-phase synchrony $y^{(1)}= y^{(2)} = y_0(x,t;(1-\gamma)^{-1}\omega)$, we add a perturbation $\delta y\ui$ and substitute back into Eq.~\eqref{eq:fbtwo}. The perturbations  $\delta y\ui$ are governed by the linear equations
$\dy\ui_t-\gamma\dy\uj_t = \mathcal{L}[y\ui;y_0]$, 
where the right-hand side is given by $ -y_{xxxx}+2[(-F+\alpha y_{0xx}^2)y_x+2\alpha y_{0x}y_{0xx}y_{xx}]_x-(-F+3\alpha y_{0xx}^2)y_{xx}$. 
By linearity, the amplitude difference, $\delta y^{-} = \dy\ua-\dy\ub$, satisfies
	$(1+\gamma)\delta y^{-}_t = \mathcal{L}[\delta y^{-}; y_0]$. Expressing in terms of the phase coordinate $\theta$, where $\theta_t = (1-\gamma)^{-1}\omega$, we arrive at the following linear equation about the in-phase state, 
\begin{align}\label{eq:inL}
	\frac{1+\gamma}{1-\gamma}\omega\delta y^{-}_\theta = \mathcal{L}[\delta y^{-}; y_0(x,\theta)].
\end{align}
Similarly for the anti-phase state, we  can derive an equation for the sum of perturbations, $\delta y^{+} = \dy\ua+\dy\ub$, 
\begin{align}\label{eq:antiL}
	\frac{1-\gamma}{1+\gamma}\omega\dy^+_\theta = \mathcal{L}[\dy^+; y_0(x,\theta)].
\end{align}
The solutions $\dy^{\pm}$ are of the form $\dy^\pm =\dy^\pm_0 e^{\mu t}$, where $\dy^\pm_0$ is periodic and $\mu$ is the growth rate. We compute the associated  Floquet multipliers $\rho\sim e^{\mu T}$ \cite{Floquet1883}, by numerically integrating Eq.~\eqref{eq:inL} and \eqref{eq:antiL} over one period $T$. For $|\rho|<1$, $=1$ or $>1$, 
 the corresponding synchronized state is stable, marginally stable, or unstable, respectively. 
In  Fig.~\ref{fig:synctime}(b,c), we plot $|\rho|$ versus $\gamma$ for in-phase and anti-phase synchrony, respectively, and for $F =42, 46$, $48$ and $49$. For $F=46$, $\gamma =0.1$, one has $|\rho|< 1$ for both modes, consistently with cases \rom{1} and \rom{2} of Fig.~\ref{fig:modes}. For $F = 49$, $\gamma = 0.1$ and for $F=48$, $\gamma = 0.01$, in-phase synchrony is stable while anti-phase is not,  as in cases \rom{3} and \rom{4}. 
Further, the Floquet multipliers are consistent with all numerical predictions of in-phase and anti-phase stability reported in Fig.~\ref{fig:basins} (see SI and Fig.~S3 \cite{SI}). Although this analysis does not shed light on the stability of the states with nontrivial phase lags, it does  show that  these states occur at values of $F$ and $\gamma$ for which anti-phase synchrony is unstable.

\begin{figure*}[!t]
	\centering
	\includegraphics[scale=1]{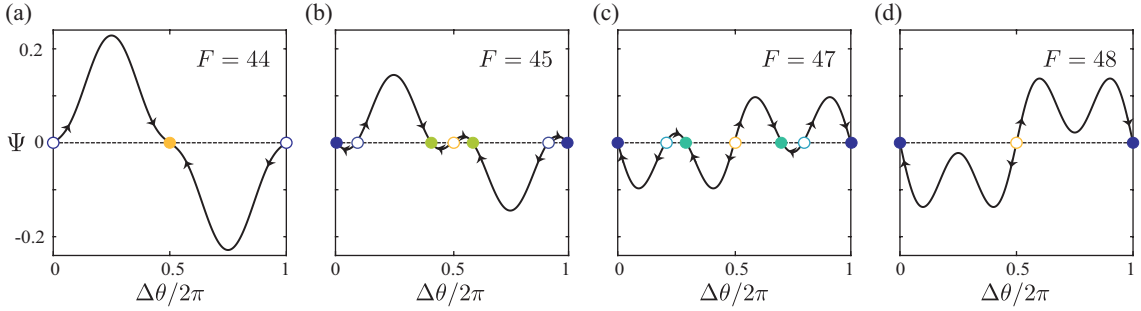}
	\caption{\footnotesize
		Phase function $\Psi$ versus phase difference $\Delta \theta$ at (a) $F = 44$, (b) 45, (c) 47 and (d) 48. The solid and hollow dots correspond to stable and unstable fixed points, respectively. The color scheme for $\Delta\theta$  uses the color map in FIG.~\ref{fig:basins}.  }
	\label{fig:adjoint}
\end{figure*} 
We posit that when synchrony is stable, the Floquet multiplier $\rho$ indicates the time it takes to synchronize. This time scales as $-1/\mu \sim 1/\ln|\rho|^{-1}$ ($\mu<0$ for stable synchrony).
In Fig.~\ref{fig:synctime}(a), we verify this finding numerically by calculating the number of periods until synchrony is reached in the nonlinear simulations and plotting versus $1/\ln|\rho|^{-1}$. 
Small $|\rho|$ indicates fast synchronization. Interestingly, closer filaments with stronger hydrodynamic coupling (larger $\gamma$) do not always exhibit more efficient synchronization; while anti-phase synchronization is always achieved faster as the interfilamentous distance gets smaller (Fig.~\ref{fig:synctime}c),  in-phase synchronization is most efficient at intermediate coupling (Fig.~\ref{fig:synctime}b).

To investigate the stability of all synchronized states, including those with nontrivial phase lag, we derive an evolution equation for the phase difference $\Delta \theta$ in the case of weak coupling. We use  the Kuramoto phase reduction approach
assuming that the dynamics of each filament asymptotically follows the single filament solution $y_0(x,t, \theta^{(i)})$, albeit at a different phase~\cite{Kawamura2018}. 

We rewrite Eq.~\eqref{eq:fbtwo} in terms of the the eigenfunction $u_0(x,\theta\ui) = \partial y_0/\partial\theta|_{\theta = \theta\ui}$  associated with the zero eigenvalue of the linear operator $(\mathcal{L} - \omega\partial/\partial \theta)$, and introduce the normalized adjoint function $\hat{u}_0(x,\theta)$ (see SI \cite{SI}). We project the resulting equation onto the single filament solution $u_0$ to obtain $\theta\ui_t =\omega[1+\gamma H(\theta\ui,\theta\uj)]$, where $H(\theta\ui,\theta\uj) = \int_0^1 \hat{u}_0(x,\theta\ui)u_0(x,\theta\uj) dx$ is the phase coupling function. We average $H$ over one cycle along the line $\theta\ub = \theta\ua+\Delta\theta$. The averaged $\bar{H}(\Delta\theta)$ depends only  on the phase difference $\Delta\theta$. We arrive at the evolution equation
\begin{align}\label{eq:phaseD}
 \Delta\theta_t =\gamma\omega \Psi(\Delta\theta),
 \end{align}
 where $\Psi(\Delta\theta) = \bar{H}(-\Delta\theta)-\bar{H}(\Delta\theta)$. Clearly, $\Psi(\Delta\theta)=0$ corresponds to equilibrium solutions of~\eqref{eq:phaseD} for which the two filaments are in synchrony. The sign of $\partial \Psi/\partial (\Delta\theta)$ at these synchronized states indicates their stability:
 $\partial \Psi/\partial (\Delta\theta)$ is positive for unstable states and vice-versa.

In Fig.~\ref{fig:adjoint}, we plot $\Psi$ as a function of $\Delta\theta$ for  $F=44$, $45$, $47$ and $48$. These plots reveal two types of bifurcations underlying the transitions displayed in Fig.~\ref{fig:basins}(a). At $F=44$, in-phase synchrony is unstable and anti-phase synchrony is stable. 
Two supercritical pitchfork bifurcations take place as $F$ increases from 44 to 45, by which the anti-phase synchrony becomes unstable and two stable equilibria appear at a nontrivial phase lag, and simultaneously, the in-phase synchrony becomes stable and two unstable equilibria appear at a nontrivial phase lag.
The location of these equilibria changes with $F$.
As $F$ increases from $47$ to $48$, two saddle-node bifurcations occur and the nontrivial equilibria vanish leaving stable in-phase and unstable anti-phase synchrony.

\begin{figure}[!t]
	\centering
	\includegraphics[scale=1]{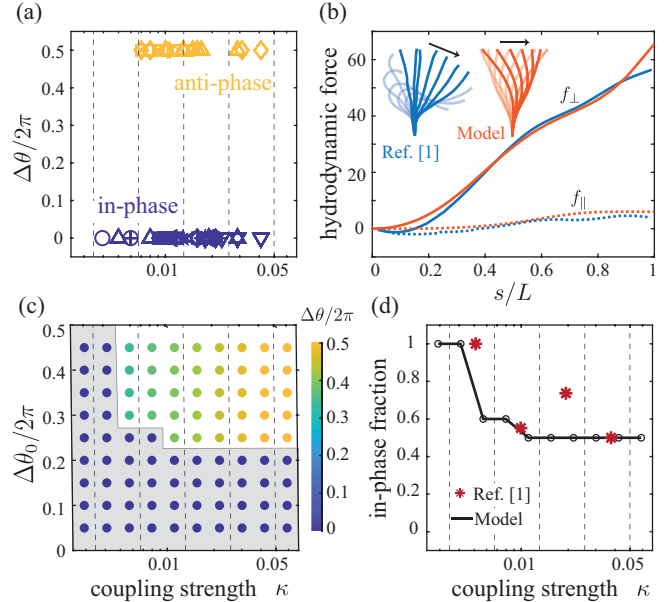}
	\caption{\footnotesize (a) Experimental data
from~\cite[Fig. 5A]{Brumley2014} represented in terms of steady-state phase difference  $\Delta \theta/2\pi$ and coupling strength $\kappa$.
(b) Normal (solid line) and tangential (dashed line) hydrodynamic forces based on the filament model with $F=50$ (orange) compared to flagellar forces (blue)~\cite[Fig.~4B]{Brumley2014}. Insets show snapshots of the filament waveform (orange) and flagellar waveform (blue)~\cite[Fig.~4A]{Brumley2014}. (c) Model prediction of synchronization modes for $F = 50$ as a function of coupling strength $\kappa$. The grey region shows the basin of attraction of in-phase synchrony.   (d) Fraction of in-phase basin of attraction to all initial phase differences;  model results from (c) are shown in solid black line, experimental data from (a) are superimposed as red stars.}
	\label{fig:experiment}
\end{figure}

Lastly, we examine key features observed in experiments with isolated flagella of {\em Volvox carteri}~\cite{Brumley2014} in light of our filament model. Results taken from~\cite[Fig. 5A]{Brumley2014} are shown in Fig.~\ref{fig:experiment}(a); in-phase and anti-phase synchronous beating were reported for a range of coupling strength $\kappa$, where $\kappa$ is defined in terms of the time it takes for the two flagella to synchronize. Based on our Floquet analysis,  $\kappa$ can be expressed as $\kappa = -\mu/(2\pi\omega)$, where $-\mu=- \omega  \ln\rho/(2\pi)$ and $\rho$ depends on $F$ and $\gamma$. We arrive at $\kappa = \ln(1/\rho)/(4\pi^2)$, which for fixed $F$ defines a map from $\gamma$ to $\kappa$ (see SI \cite{SI}).

We first match the filament active force $F$ and frequency of oscillation $\omega/2\pi$ to those of the flagella. The measured flagellar frequency was about 30 \si{\hertz} and  total force about 50 \si{\pico\newton} \cite{Brumley2014}. Using flagellar length $L = 20$  \si{\micro\meter} and bending rigidity $B = 4\times 10^{-22}$ \si{Nm^2},  the dimensionless counterparts are $\omega/2\pi = 48$ and $F = 50$. For active force $F=50$ in our model, the resulting filament frequency $\omega/2\pi = 46$ (Fig. S1,d) is close to that of the flagellar beat, and the distribution of forces along the filament and flagellum are also similar (Fig.\ref{fig:experiment}b).

In Fig.~\ref{fig:experiment}(c), we show the synchronization modes of a filament pair for the range of $\kappa$ reported in~\cite{Brumley2014}. At small $\kappa$, only in-phase synchrony is stable. As $\kappa$ increases, both in- and anti-phase synchrony are stable, consistent with Fig.~\ref{fig:basins}.  The fraction of all initial phase differences (grey region in Fig.~\ref{fig:experiment}c) that lead to in-phase synchrony is shown in Fig.~\ref{fig:experiment}(d): fraction value 1 indicates that only in-phase synchrony is stable while 0.5 means bistable in- and anti-phase synchrony with equal-size basins of attraction.  To compare to Fig.~\ref{fig:experiment}(a), we interpret the experimental data as random samples from the phase space in~Fig.~\ref{fig:experiment}(c), we divide $\kappa$ evenly in log-space into four ranges and we count the instances of in-phase synchrony in each range. The fraction of in-phase to total number of data points in each range are shown as red dots in Fig.~\ref{fig:experiment}(d). The results agree remarkably well with the filament model.

These findings could be instrumental for deciphering the biophysical and biochemical mechanisms underlying transitions in flagellar synchrony~\cite{Brumley2014,Wan2014,Leptos2013}.  Such transitions could be triggered mechanically, say by random disturbances causing a shift between bistable modes, or physiologically by modifying either the intensity of the filament activity or interfilamentous coupling. The latter, in addition to hydrodynamics, could be due to basal connections between the flagella in the cell surface~\cite{Wan2016, Liu2018,GuoArxiv}. These considerations, as well as extensions to arrays of microfilaments, will be treated in future works.

\newpage

\end{document}